\documentclass[review]{elsarticle}

\usepackage{lineno,hyperref}
\usepackage{latexsym,amssymb,amsmath}
\usepackage{graphicx}
\usepackage{color}
\journal{Journal of \LaTeX\ Templates}

\biboptions{authoryear}

\begin{document}

\begin{frontmatter}

\title{Excitation of flare-induced waves in coronal loops and the effects of radiative cooling}

\author{Elena Provornikova}
\address{University Corporation for Atmospheric Research (UCAR), Boulder, CO, 80307, USA}
\address{Naval Research Laboratory, Washington, DC 20375, USA}

\author{Leon Ofman\fnref{myfootnote}}  
\author{Tongjiang Wang}
\address{Department of Physics, Catholic University of America, Washington, DC 20064, USA}
\address{NASA Goddard Space Flight Center, Code 671, Greenbelt, MD 20771, USA}
\fntext[myfootnote]{Visiting, Department of Geosciences, Tel Aviv University, Tel Aviv, Israel}




\begin{abstract}
EUV imaging observations from several space missions (SOHO/EIT, TRACE, and SDO/AIA) have revealed a presence of propagating intensity disturbances in solar coronal loops. These disturbances are typically interpreted as slow magnetoacoustic waves. Recent spectroscopic observations with Hinode/EIS of active region loops, however, revealed that the propagating intensity disturbances are associated with intermittent plasma upflows (or jets) at the footpoints which are presumably generated by magnetic reconnection. For this reason, whether these disturbances are waves or periodic flows is still being studied. This study is aimed at understanding the physical properties of observed disturbances by investigating the excitation of waves by hot plasma injections from below and the evolution of flows and wave propagation along the loop. We expand our previous studies based on isothermal 3D MHD models of active region to a more realistic model that includes full energy equation accounting for effects of radiative losses. Computations are initialized with an equilibrium state of a model active region using potential (dipole) magnetic field, gravitationally stratified density and temperature obtained from polytropic equation of state. We model an impulsive injection of hot plasma into the steady plasma outflow along the loops of different temperature, warm ($\sim$1 MK) and hot ($\sim$6 MK). The simulations show that hot jets launched at the coronal base excite slow magnetoacoustic waves that propagate along the loops to the high corona, while the injected hot flows decelerates rapidly with heights. The simulated results support that the observed coronal disturbances are mainly the wave features. We also find that the effect of radiative cooling on the damping of slow-mode waves in 1-6 MK coronal loops is small, in agreement with the previous conclusion based on 1D MHD models. 
\end{abstract}

\begin{keyword}
corona; magnetohydrodynamics (MHD); waves; oscillations 
\end{keyword}

\end{frontmatter}

\section{Introduction}
Waves in the EUV are commonly observed phenomena in the solar corona thanks to the high resolution, high cadence multi wavelength observations by SDO/AIA. The observations of EUV waves in the corona were recently reviewed by \citet{LO14} and \citet{wan16}. Since first observation of propagating intensity disturbances (PDs) in white light polarized brightness along coronal plumes \citep[e.g.,][]{Ofm97}, such phenomena were commonly detected in plumes (see review by \citet{ban16}) and in large, quiescent coronal loops by different space instruments including SOHO/EIT \citep{DG98}, TRACE \citep{DHA00}, Hinode/EIS \citep{wan09, kit10}, SDO/AIA \citep{kri12a, kri12b, uri13}. PDs often appear in the loops near the edge of active regions and seen as small, by few percent, amplitude changes in EUV intensity, and were found to propagate at approximately the local sound speed.  PDs observed in coronal loops are generally quasi-periodic with periods of the order of a few minutes ranging in 2.5 - 9 min \citep{DeM09}. The phenomena of PDs observed both in coronal plumes and in large fanlike coronal loops were interpreted as slow magnetosonic waves \citep{OND99,Rob99,Nak00}. In addition, Doppler shift oscillations were first observed with SOHO/SUMER in hot ($>$6 MK) flaring loops \citep{Wan02}, and were interpreted as standing slow-mode waves \citep{OW02}. The trigger of hot loop oscillations was found to be associated with high-speed (200-300 km/s) flow pulses, which were likely produced by small (or micro-) flares at the footpoints \citep{wan05}.

Observations of coronal waves and loop oscillations are important for a new rapidly developing research area, coronal seismology \citep[see the review by][ and references therein]{Nak16}, which aims to infer physical parameters of the solar atmosphere from observations of coronal waves. Methods of coronal seismology may provide information on the corona magnetic field e.g., \citet{NO01,NV05} and dissipation coefficients in coronal plasma \citep[e.g.,][]{Wan15}. Slow magnetoacoustic waves can provide a significant contribution to the heating of coronal loops \citep{TN01}. In particular, PDs are believed to be driven below the corona making this phenomenon important for studies of the connection between the corona and chromosphere/transition region \citep{DeM02}. Recently, the origin of magnetoacoustic waves observed above sunspots was identified from $p$-mode helioseismic waves traveling upward through different atmospheric levels \citep{zha16}. By using seismology techniques, the sunspot waves have been applied to determine the coronal magnetic field strength \citep{jes16}.

However, the interpretation of PDs in non-sunspot loops is still controversial. The main reason is that these PDs were discovered to be associated with high-speed ($\sim$50 km/s) outflows with the quasi-periodic features, and so have been alternatively interpreted as periodic upflows (or jets) by several authors \citep{dep10, tian11}. Recognition of the true nature of PDs is crucial to their seismological application and to understand their role in coronal heating and mass supplies. Some 3D MHD simulations have shown that quasi-periodic outflows injected at the footpoints of coronal loops inevitably generate slow magnetoacoustic waves propagating upwards along the loop \citep{OWD12,WOD13}. The simulated flows decelerate with height rapidly, suggesting that the observed PDs at higher corona ($r \gtrsim 1.2 R_{S}$) are due to slow magnetosonic wave features.

Despite of numerous works on observations and theory of propagating slow magnetoacoustic waves in coronal loops, it is still not completely understood what are the driving mechanisms and what processes cause a rapid damping of slow waves in the corona. 
Amplitudes of waves show rapid decay as they travel outward along the loop with the decay time of the order of minutes. Recent studies indicate that observed wave damping is frequency-dependent \citep{KPrasad14,Mandal16}. The observed propagation velocities and damping lengths of slow waves are also temperature-dependent \citep{kri12b, uri13}. Several theoretical studies showed that dissipative processes of thermal conduction, compressive viscosity, radiative cooling and effect of magnetic field divergence can lead to a decay of amplitude of propagating slow magnetoacoustic waves. High thermal conduction in hot coronal loops with temperature $\sim$6 MK results in rapid damping of slow waves on timescales on the order of wave period \citep{OW02}. Other studies by \citet{PKS94,Nak00,DeM03,Kli04} also found that thermal conduction is the dominant damping mechanism for slow waves in the coronal loops while other processes and effects have smaller contribution. However thermal conduction process alone does not account for the observed damping \citep{Sig07}. 

Coronal radiative losses are small compared to the energy losses due to thermal conduction. For typical quiescent corona ($T=1\,MK,\, n = 10^9\, cm^{-3}$) characteristic time of radiative cooling is about 30 minutes, while for thermal conduction the timescale is in the order of few seconds. Several studies showed that in certain plasma parameters cooling can significantly affect propagation of magnetoacoustic waves in coronal loops. \citet{AlGRu14} showed that under certain assumptions, the period of standing slow waves increases due to the cooling. Cooling affects waves differently in the loops of different temperatures. In cooler loops (below $1\,MK$) cooling leads to the wave amplification increasing its amplitude. In very hot  6 MK loops cooling causes an enhancement of wave damping (which is mainly due to thermal conduction). 
\citet{BradErd08} considered the effects of radiative cooling on damping of standing slow waves in the model loop that extends from the chromosphere to the corona through the transition region. It is expected that in the transition region, where plasma is denser than in the corona and contains some of the most strongly emitting ions, wave energy would efficiently radiate away. They showed that radiative emission arising from a non-equilibrium ionization balance reduces the damping timescale in comparison to the equilibrium case by up to 10\%. 

We are motivated by the association of observed PDs with recurrent outflows at the coronal base, and their temperature dependence in propagation speed and decay length, and aim to determine a role of different dissipation processes in damping of magnetoacoustic waves in loops with different plasma parameters, e.g. cooler quiescent loops and hotter flaring loops, in a three-dimensional (3D) geometry. In this work we study the flow-generated waves for probing the physical properties of PDs and 
the effects of optically thin radiative cooling on the simulated PDs in coronal loops. In our model PDs are driven by the hot, pulsed upflows at the loop footpoints. Hot plasma upflows or jets could be originated in nanoflares at the base or below the corona. We expand the previous isothermal studies by \citet{OWD12} and \citet{WOD13} of propagating slow waves in coronal loops by including a full energy equation in the 3D MHD model. This improvement allows us to study thermodynamic evolution of PDs such as due to radiative cooling and thermal conduction in a 3D geometry. Focusing on the effects of radiative cooling in this paper, we leave the study of thermal conduction effects that are more challenging for the numerical implementation for the future study.

\section{MHD model for waves in coronal loops}\label{model}

To model plasma  dynamics in a 3D dipole active region we solve the time-dependent MHD equations including effects of gravity, resistivity and cooling due to radiative losses from optically thin plasma. The MHD equations are written in the dimensionless form:

\begin{eqnarray}
\frac{\partial \rho}{\partial t} + \nabla \left( \rho \mathbf{V} \right) = 0 \\
\frac{\partial \left( \rho \mathbf{V} \right)}{\partial t} + \nabla \cdot \left[ \rho \mathbf{V} \mathbf{V} +  \left( Eu \, p + \frac{\mathbf{B} \cdot \mathbf{B}}{2} \right) \mathbf{I} - \mathbf{B} \mathbf{B} \right] = - \frac{1}{Fr} \rho  \mathbf{F_{grav}}  \\
\frac{\partial \rho E}{\partial t} + \nabla \left[ \mathbf{V} \left(\rho E + Eu\,p + \frac{\mathbf{B} \cdot \mathbf{B}}{2} \right) -  \mathbf{B}(\mathbf{B}\cdot \mathbf{V}) + \frac{1}{S} \nabla \times \mathbf{B} \times \mathbf{B}  \right] =\nonumber \\ \frac{1}{Fr} \rho \mathbf{F_{grav}}\cdot \mathbf{V} - Q_{rad} + H \label{E:eq}\\
\frac{\partial \mathbf{B}}{\partial t} = \nabla \times \left( \mathbf{V} \times \mathbf{B} \right) + \frac{1}{S}\nabla^2 \mathbf{B}
\end{eqnarray}

Here $E$ is the total energy density \begin{equation} E = \frac{p}{(\gamma -1) \rho} + \frac{\mathbf{V}^2}{2} + \frac{\mathbf{B}^2}{2 \rho}.\end{equation} Temperature in the quiescent corona changes gradually with height resulting in nearly isothermal structure \citep{Asc05}. For simplicity in the energy equation we neglect the explicit heating terms and assume polytropic index $\gamma = 1.05$ to model the effects of heating of coronal plasma. The implied heating may also reduce the magnitude of the temperature decrease due to the radiative cooling. This empirical value of $\gamma$ produces gradually decreasing temperature with height profile. This value is commonly used in global MHD models of the corona and solar wind \citep{Mik99,Coh07} and in global multi-fluid models \citep{Ofm15}. Using $\gamma = 5/3$ with explicit empirically obtained heating terms, radiative, and thermal conductive losses is an alternative approach, where the modeling of the heating is more direct - is the subject of a future study. In the above equations the gravity term $\mathbf{F_{grav}}$ is modeled as $\mathbf{F_{grav}} = \frac{\rho}{(R_S + z - z_{min})^2/(0.1R_S)^2}$, assuming small hight of the active region compared to the solar radius, $R_S$. 

Governing dimensionless parameters appeared in the equations after normalization are Euler number $Eu = \frac{\beta}{2} = \frac{4 \pi p_0}{B_0^2}$ characterizing the ratio of plasma thermal pressure and magnetic pressure (where $\beta\ll 1$ in a typical coronal active region),  Froude number $Fr =  \frac{V_0^2 L_0}{G M_S}$; Lundquist number $S = \frac{L_0 V_0}{\eta} $. In the simulations we use the following characteristic parameters: distance scale $L_0=0.1\, R_S$, magnetic field strength $B_0 = 100\,G$, temperature $T_0 = 1\, MK$, number density $n_0=1.38 \times 10^9\, cm^{-3}$, Alfv\'en speed $V_{A0}=5872\, km/s$, Alfv\'en time $\tau_A = 12\,s$, isothermal sound speed $c_s=128\,km/s$, gravitational scale height $H_0=60\,Mm$. 

\begin{figure}
\includegraphics[scale=0.35]{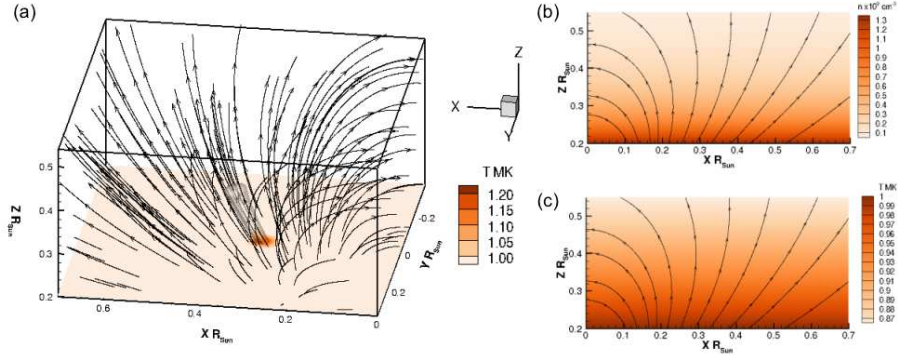}
\caption{Initial conditions for the 3D MHD simulations of fan-like loops. (a) Dipole magnetic field for the model active region. The area with higher temperature at bottom boundary is where plasma outflow and hot jets are initiated. (b) and (c) Gravitationally stratified density and polytropic temperature with $\gamma=1.05$ in the x-z plane of a 3D domain, respectively. \label{fig1}}
\end{figure}

The term $Q_{rad}$ in Equation~(\ref{E:eq}) expresses the rate of losses of thermal energy due to optically thin radiative cooling and has the form 
 \begin{equation} Q_{rad} = C_{rad} n_e^2 \Lambda (T_e), \end{equation} where $\Lambda (T_e)$ is the radiative cooling function depending on the electron temperature, $n_e$ is the electron number density and $C_{rad} = c_A^3 m_p/(n_0 L)$ is a normalization constant. In our one-temperature model we assume equal temperatures for electrons and ions (considering protons only) $T_e=T_i=T$ and plasma is quasi-neutral $n_e=n_i = n$. For plasma in temperature range $10^{6}<T<10^{7}$ the radiative cooling function can be expressed in a power law form \citep{RTV78} 
\begin{equation}\Lambda (T_e) =10^{-18.66} \frac{1}{T^{1/2}},
\end{equation}
where  $T$ is in degrees Kelvin and $\Lambda (T_e)$ is in $ergs\,cm^3\, s^{-1}$. Heating function $H = n_{t=0}^2 \Lambda(T_{t=0})$ in the energy equation is to balance radiative losses in the initial state of the system. Here $n_{t=0}=n(t=0,x,y,z)$ and $T_{t=0}=T(t=0,x,y,z)$ are initial density and temperature in a 3D domain. The initial density in polytropic atmosphere is given by 
\begin{equation} 
\frac{\rho}{\rho_0}=\left[ 1+\frac{(\gamma -1)}{\gamma H} \left(\frac{1}{10+z-z_{min}} - \frac{1}{10}\right)\right]^{1/(\gamma -1)}
\end{equation}
where  $H = 0.2 kT_0R_S/(m_pGM)$ is the normalized gravitational scale height, $M_S$ and $R_S$ is the mass and raduis of the Sun, $k_B$ is Boltzmann`s constant, $T_0$ is the temperature, $G$ is the universal gravitational constant and $m_p$ is the proton mass. Temperature is calculated from the polytropic relation as $T/T_0 = (\rho/\rho_0)^{\gamma -1}$.
Potential dipole magnetic field with gravitationally stratified density and polytropic temperature at the initial state of the system are shown on Figure~\ref{fig1}. To model fan-like loops that appear as open field at low heights, we take half of the domain of a bipolar field. The equations are solved in a 3D computational domain $(0,\,7)\times (-3.5,\,3.5) \times (2,\,5.5)$ in normalized distance units with uniform grid, typically, $258^3$. This resolution was found to be adequate for the model by applying a convergence test. The numerical method used is the modified Lax-Wendroff method with fourth-order stabilization term \citep[e.g.][]{OT02}. We consider propagation of waves in a fan-like warm (1-2 MK) loops as well as in closed hot (6 MK) flare loops. 

Similar to the isothermal study by \citet{WOD13}, we model steady plasma upflow along the magnetic field lines and impulsively inject hot material into the upflow at the bottom boundary (Figure \ref{fig1}). The boundary condition for plasma velocity at $z=z_{min}$ in the area $r \leq 2 \omega_0$ is the following:
\begin{equation} \mathbf{V} = V_0(x,y,z=z_{min},t)  \mathbf{B}/|B|,
\end{equation}
where 
\begin{equation}
V_0(x,y,z=z_{min},t)  = V_{A0} A_v(t) \exp\left[ -\left( \frac{r}{\omega_0} \right)^4 \right].
\end{equation}
Here $r = \left[ \left(x-x_0\right)^2 + \left(y - y_0\right)^2\right]^{1/2}$. The parameters are $x_0 =3.0$, $y_0 =0$ and $\omega_0=0.12$. The function $A_v(t)$ sets a single pulse in the constant background:
\begin{eqnarray}
A_v(t) =  \left\{ 
\begin{array}{ll} 
A_0 & (0<t<t_1 \,\, or \,\,t_2<t<t_{max})\\
A_0 + \frac{1}{2}A_0 \left[1 - \cos \frac{2 \pi (t-t_1)}{P} \right]& (t_1 \leq t \leq t_2)\\
\end{array} 
\right. 
\end{eqnarray} 
where the parameters $P = t_2 - t_1=30\tau_A$, $t_1=170\tau_A$, $t_2 = 200\tau_A$, $t_{max}=400\tau_A$ and $A_0 = 0.01$. The amplitude of the injected pulse is $0.02$ which is subsonic.
Similar condition is set for the plasma temperature:
\begin{equation} T(x,y,z=z_{zmin},t) = T_0 + T_0 A_T(t) \exp\left[ -\left( \frac{r}{\omega_0} \right)^4 \right],
\end{equation}
where
\begin{eqnarray}
A_T(t) =  \left\{ 
\begin{array}{ll} 
T_c & (0<t<t_1 \,\, or \,\,t_2<t<t_{max}),\\
T_c + \frac{1}{2}T_c \left[1 - cos \frac{2 \pi (t-t_1)}{P} \right]& (t_1 \leq t \leq t_2),\\
\end{array} 
\right. 
\end{eqnarray} 
where $T_c = 0.5$. For the quiet corona of $T = 1 \,MK$ with magnetic field $B = 100 \, G$ and density $n = 1.38 \times 10^9 \, cm^{-3}$ such conditions determine a steady plasma upflow along the loop with the velocity $60\, km/s$ and temperature $T = 1.5 \,MK$ at $z = z_{min}$ and a short-period subsonic injection of plasma with the velocity amplitude $120 \, km/s$ and temperature $T = 2\, MK$.

\section{Results and Discussion}

\begin{figure}
\includegraphics[scale=0.35]{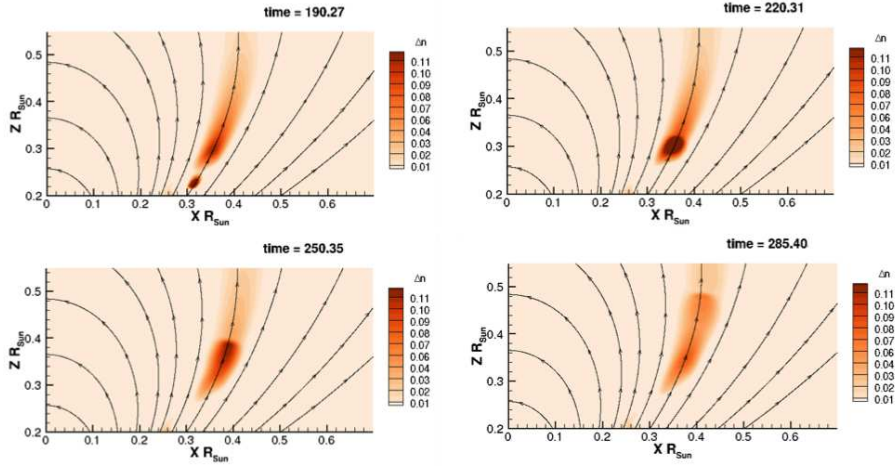}
\caption{ Snapshots of density difference $\Delta n = n - n_{t=0}$ in the $x-z$ plane at $y=0$ at different times showing propagation of the disturbance initiated by a hot jet. \label{fig2}}
\end{figure}

Figure \ref{fig2} shows snapshots of density difference $\Delta n = n - n_{t=0}$ in $x$-$z$ plane at different times after the steady outflow is set up and hot velocity pulse is initiated. Such short-time hot plasma upflow (or jet) could be produced by reconnection at the loop footpoints in active region. The pulse generates a slow magnetoacoustic wave propagating along the loop. Propagating density disturbance associated with the slow wave is shown on Figure  \ref{fig2}.

\begin{figure}
\includegraphics[scale=0.35]{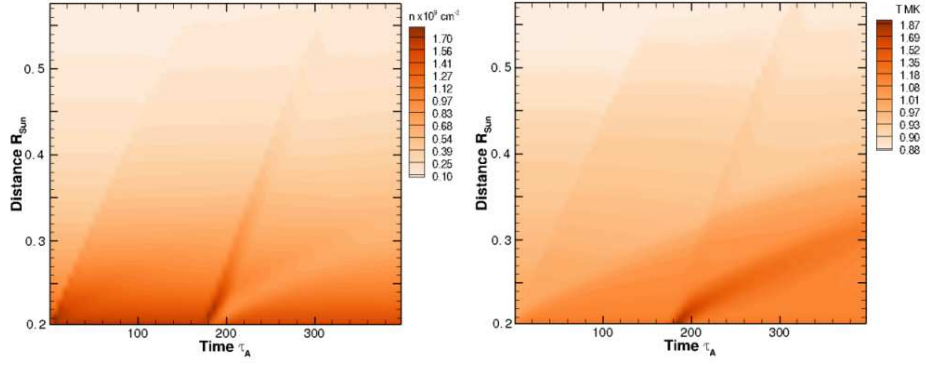}
\caption{Plasma density and temperature along the loop as a function of time showing signatures of flows and slow magnetoacoustic waves. \label{fig3}}
\end{figure}

Figure \ref{fig3} shows a time-distance diagram of density and temperature for a cut along the magnetic loop at the center of the area $r \leq 2 \omega_0$ demonstrating a propagation of a slow wave generated by the jet at $t =170\tau_A$. Behind the wave front, injected hot plasma material expands and propagates with the speed of a steady outflow in the loop which is noticeably smaller than the wave propagation speed. The average speed of the wave produced by the hot pulse can be estimated as $160\,km/s$ while the average speed of the injected hot plasma carried by the steady outflow is approximately $30 km/s$. The latter is much smaller than the injection speed of both steady outflow $60 \,km/s$ and a pulse $120\,km/s$ indicating a deceleration of flows with height. Injection of a hot pulse produces a local expansion in plasma with associated low density
which is seen on the Figure \ref{fig3} (left).

\begin{figure}
\includegraphics[scale=0.8]{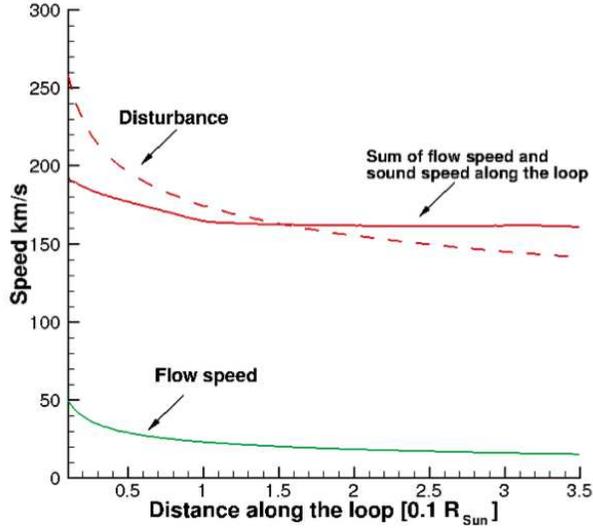}
\caption{Profiles of the velocity of the steady background outflow $V_{bg}$ (green curve),  estimated velocity of slow wave propagation as $V_{bg} + c_s$ (solid red curve) and velocity of the wave calculated from simulations (dashed red curve). Profiles are calculated along the magnetic line originating at $(x_0=3,\, y_0=0)$, the center of the region where hot plasma jet is applied. \label{fig4}}
\end{figure}

A comparison of various speed profiles is shown on Figure \ref{fig4}. Green curve shows the speed of the background plasma outflow along the loop $V_{bg}$ before the injection of the hot pulse. This curve was obtained by extracting the plasma speed along the magnetic line at $t = 170\tau_A$. The red solid curve shows a profile of the speed of the slow magnetoacoustic wave in the presence of the steady flow $V_{bg} + c_s$, where $V_{bg}$ is the speed of the outflow (shown by green curve) and $c_s$ is the sound speed. The sound speed $c_s$ is calculated using the profile of temperature in the steady outflow at $t=170 \tau_A$. The speed of the disturbance generated by the hot jet is shown by dashed red curve. It was calculated from differences in locations of a PD in 1D cuts along the magnetic loop at subsequent moments of time as $V_{PD} = (s_{t2} - s_{t1})/(t2-t1)$. 

At lower heights the speed exceeds the value of $(V_{bg} + c_s)$ due to the presence of the pulse with the velocity amplitude of  $120\, km/s$. But in general the profile of the disturbance propagation speed agrees well with the estimation $V_{bg} + c_s$. The comparison of speed profiles suggests that small-scale hot plasma upflows produce disturbances propagating along the magnetic loop which clearly show signatures of slow magnetoacoustic waves. While waves can propagate to high altitudes, the signatures of high speed plasma upflows are only seen at lower heights up to 10 Mm. These results are consistent with the previous study by \citet{WOD13}.

\begin{figure}
\includegraphics[scale=0.38]{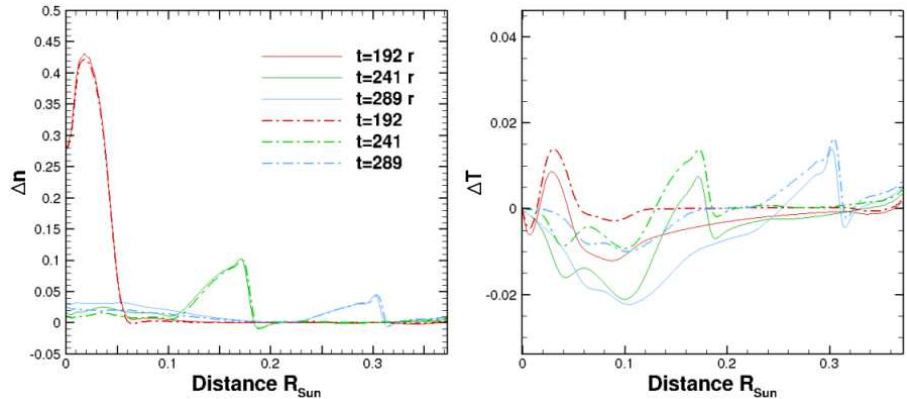}
\caption{ Difference in plasma density $\Delta n = n - n_{t=0}$ ({\it left}) and temperature $\Delta T = T - T_{t=0}$ ({\it right}) along the loop for different times. Simulation with radiative losses (solid curves) are compared with simulations when radiative losses are neglected (dash-dotted curves). X-axis show the distance along the loop starting from the bottom boundary of the domain. \label{fig5}}
\end{figure}

To explore the effects of radiative cooling on propagating slow magnetoacoustic waves we compare two simulations, with and without the radiative cooling term (and balancing heating term) in the energy equation. Figure~\ref{fig5} shows the profiles of $\Delta n$ and $\Delta T$ along the magnetic loop at different times for two different simulations. It is seen that the effect of radiative losses is small and does not significantly change the amplitude of the disturbance and its propagation speed. The explanation can be given if we compare a characteristic cooling time due to radiative losses and a timescale of wave propagation in a fan-like loop.
The propagation time of the slow wave in a loop with the length of $200\, Mm$ with the velocity of $\sim 170\,km/s$ is $\tau_{w} \approx 10^3 \,s$ which is smaller or comparable to the radiative cooling timescale $\tau_{rad} \approx 3 \cdot 10^3 \,s$. The radiative cooling timescale can be calculated as $\tau_{rad} = \frac{3 k_b T_e}{n_e \Lambda(T_e)}$ \citep{Asc05,AT08}. For hot flare loops with $T \sim 6 MK$ the radiative cooling  timescale is even larger $\tau_{rad} \approx 2\cdot 10^4\, s$. So one would expect that radiative cooling would not result in the effective damping of waves in hot flare loops if propagation time is of the order of the radiation timescale. We performed simulations with hot $6\, MK$ loops \citep{OWD12} and confirmed that the effect of radiative cooling appears to be negligible. 

\section{Conclusion and Future Work}

Observations of propagating disturbances in coronal loops in EUV show a rapid decrease of their amplitudes with height with damping timescales on the order of minutes. While a number of previous works, mostly restricted to 1D models,  suggested that the dominant damping mechanism for PDs in the corona is thermal conduction, that process alone can not account for the observed damping timescales. To quantify a contribution of various dissipation processes, such as thermal conduction, radiative cooling and compressive viscosity, it is required to study excitation and propagation of the disturbances in a 3D model of active region with realistic profiles of the coronal plasma parameters. In this work we focused on the effects of optically thin radiative losses on propagating slow magnetoacoustic waves in coronal loops induced by hot plasma inflows at the base of the corona. We leave a study of thermal conduction effects that are more challenging for the numerical implementation for the future work. 

The presented model expands previous isothermal studies of damping of injected slow magnetosonic waves in a model active region developed by \citet{OWD12} and \citet{WOD13}. We improved our 3D MHD model by including full energy equation accounting for losses of thermal energy by radiation in this particular study. The model now allows us to model more realistically the excitation of waves by hot plasma uplows originated by flares and study the thermal effects on the evolution of slow waves in coronal loops. We studied an excitation and propagation of slow magnetoacoustic waves in a 3D dipole active region in the corona with gravitationally stratified density and temperature obtained from polytropic equation of state with $\gamma = 1.05$. We modeled an impulsive small-scale injection of hot plasma into the steady plasma outflow along the loops of different background temperatures, e.g. quiescent loops ($T \sim 1\, MK$) and flare loops ($T \sim 6\, MK$). We confirm the conclusions of  \citet{OWD12} and  \citet{WOD13} that injection of flows (pulse) at the loop's footpoint inevitably excite magnetoacoustic waves propagating upward to the high corona. This supports the interpretation of observed PDs (at least at high corona) by slow-mode waves. We explored the effects of optically thin radiative cooling on propagating slow magnetoacoustic waves by comparing simulations with a without the radiative cooling term in the energy equation. Results showed that radiative cooling has a small effect on slow waves decreasing the amplitude of temperature oscillations by a few percent in the case of 1 MK loops. The effect for 6 MK loops is negligible. The propagation time of a slow wave with phase speed $\sim 170\,km/s$ in a 1 MK loop with length $200\, Mm$ is $\tau_{w} \approx 10^3 \,s$  which smaller or comparable to the radiative cooling timescale $\tau_{rad} \approx 3 \cdot 10^3 \,s$.  For hotter loops the radiative timescale is larger by the order of magnitude. These estimates show that one should expect the radiative cooling not to cause an effective damping of slow magnetoacoustic waves in the corona due to rather long timescales for losses of energy due to radiation. Our conclusion agrees with the previous 1D study by \citet{DeM04}. 

 Several previous studies \citet{OW02,DeM03} showed that thermal conduction could be an important  dissipation mechanism of slow magnetosonic waves in the coronal loops.  The present model does not include explicit effects of thermal conduction and coronal heating (often modeled by using an empirical heating function) - this extension of the model will be the focus of a future study. Including thermal conduction in the 3D MHD model will also allow us to perform parametric studies and explore damping rates for various thermal conductivities, including recently found suppressed thermal conduction in hot flaring loops \citet{Wan15}.

\citet{DeM04} showed that damping rate of slow waves in coronal loops is sensitive to the magnetic field geometry and gravitational stratification in the active region. The divergence of magnetic field can cause significant additional decrease of wave amplitude. Both of these effects are self-consistently included in our 3D MHD model but effects of magnetic field geometry are limited to the dipole structure. Implementing more realistic magnetic field structure of an active region such as potential or force-free magnetic field extrapolations from the photospheric magnetic data from SDO/HMI would make our 3D model suitable for the direct comparison with EUV observations. Hot plasma upflows could be generated by the magnetic reconnection process in the chromosphere or transition region where optically thin approximation is not applicable. To study the propagation of waves in more realistic model would require an inclusion of dense cool chromosphere and radiation transport in the model.

\section*{Acknowledgement}
LO would like to acknowledge support by NASA grant NNX16AF78G. TW was supported by the NASA Cooperative Agreement NNG11PL10A to CUA. EP acknowledge support of NASA grant NNX12AB34G and basic research funds from the Chief of Naval Research. Also EP performed a part of this work during the Postdoctoral Jack Eddy Fellowship administered by UCAR. Resources supporting this work were provided by the NASA High-End Computing (HEC) Program through the NASA Advanced Supercomputing (NAS) Division at Ames Research Center.

\section*{References}

\bibliography{ofman_klu}

\end{document}